\documentclass[english]{revtex4-1}
\usepackage[T1]{fontenc}
\usepackage[latin9]{inputenc}
\usepackage{color}
\usepackage{array}
\usepackage{float}
\usepackage{multirow}
\usepackage{amsmath}
\usepackage{stmaryrd}
\usepackage{graphicx}

\makeatletter

\providecommand{\tabularnewline}{\\}

\makeatother

\usepackage{babel}
\begin{document}

\title{Color Octet Electron Search Potential of the FCC Based e-p Colliders}

\author{Y. C. Acar}

\affiliation{Department of Electrical and Electronics Engineering, TOBB University
of Economics and Technology, Ankara, Turkey}
\email{ycacar@etu.edu.tr}

\author{U. Kaya}

\affiliation{Department of Material Science and Nanotechnology, TOBB University
of Economics and Technology, Ankara, Turkey}

\affiliation{Department of Physics, Faculty of Sciences, Ankara University, Ankara,
Turkey}
\email{ukaya@etu.edu.tr}

\author{B. B. Oner}

\affiliation{Department of Material Science and Nanotechnology, TOBB University
of Economics and Technology, Ankara, Turkey}
\email{b.oner@etu.edu.tr}

\author{S. Sultansoy}

\affiliation{Department of Material Science and Nanotechnology, TOBB University
of Economics and Technology, Ankara, Turkey}

\affiliation{ANAS Instute of Physics, Baku, Azerbaijan}
\email{ssultansoy@etu.edu.tr}

\begin{abstract}
Resonant production of color octet electron, $e_{8}$, at the FCC
based ep colliders has been analyzed. It is shown that e-FCC will
cover much a wider region of $e_{8}$ masses compared to the LHC.
Moreover, with highest electron beam energy, $e_{8}$ search potential
of the e-FCC exceeds that of FCC pp collider. If $e_{8}$ is discovered
earlier by the FCC pp collider, e-FCC will give opportunity to handle
very important additional information. For example, compositeness
scale can be probed up to hundreds TeV region.
\end{abstract}
\maketitle

\section{Introduction}

Standard Model (SM) has proven its reliability by the experimental
\textcolor{black}{verifications of its particle content in the recent
decades. SM puzzle has been completed by the discovery of Higgs boson
\cite{key-1,key-2}. However, SM seems not to be the end of the whole
story. There are still many unsolved problems that are out of the
scope of the SM and especially the large number of currently known
elementary particles becomes more of an issue. For this reason a lot
of BSM models have been proposed including extension of scalar and
fermionic sectors of SM, enlargement of SM gauge symmetry group, SUSY,
compositeness (preons \cite{key-3}), extra dimensions etc. Keeping
in mind historical development of fundamental building blocks of matter,
the search for preonic models seem to be quite natural. This development
is summarized in Table I. }

\textcolor{black}{}
\begin{table}[H]
\begin{centering}
\textcolor{black}{\caption{Historical development of fundamentality.}
}
\par\end{centering}
\centering{}\textcolor{black}{}%
\begin{tabular}{|r|c|c|c|}
\hline 
\textcolor{black}{Stages} & \textcolor{black}{1870-1930s} & \textcolor{black}{1950-1970s} & \textcolor{black}{1970-2030s}\tabularnewline
\hline 
\hline 
\textcolor{black}{Fundamental Constituent Inflation} & \textcolor{black}{Chemical Elements} & \textcolor{black}{Hadrons} & \textcolor{black}{Quarks, Leptons}\tabularnewline
\hline 
\textcolor{black}{Systematics} & \textcolor{black}{Periodic Table} & \textcolor{black}{Eight-fold way} & \textcolor{black}{Family Replication}\tabularnewline
\hline 
\textcolor{black}{Confirmed Predictions} & \textcolor{black}{New Elements} & \textcolor{black}{New Hadrons} & \textcolor{black}{BSM particles}\tabularnewline
\hline 
\textcolor{black}{Clarifying Experiments} & \textcolor{black}{Rutherford} & \textcolor{black}{SLAC-DIS} & \textcolor{black}{LHC or rather FCC?}\tabularnewline
\hline 
\textcolor{black}{Building Blocks} & \textcolor{black}{Proton, Neutron, Electron} & \textcolor{black}{Quarks} & \textcolor{black}{Preons?}\tabularnewline
\hline 
\textcolor{black}{Energy Scale} & \textcolor{black}{MeV} & \textcolor{black}{GeV} & \textcolor{black}{TeV?}\tabularnewline
\hline 
\textcolor{black}{Impact on Technology} & \textcolor{black}{Exceptional} & \textcolor{black}{Indirect} & \textcolor{black}{Exceptional }\tabularnewline
\hline 
\end{tabular}
\end{table}

\textcolor{black}{Family replication and especially SM fermion mixings
can be considered as indications of preonic structure of matter. One
of the notable results of preonic models is prediction of well-known
BSM particles (such as excited leptons and quarks, leptoquarks) and
contact interactions which are widely investigated by ATLAS and CMS.
In composite models with colored preons (see \cite{key-4} and references
therein), leptons have color octet partners, $\ell_{8}$, which are
known as leptogluons. Phenomenologically their status is similar to
excited leptons and leptoquarks. Experimentally excited leptons and
leptoquarks are considered in CMS and ATLAS experiment searches, however,
there is no direct search on leptogluons.}

\textcolor{black}{There are a number of phenomenological studies on
$\ell_{8}$ production at TeV colliders. For example, production of
leptogluons at the LHC has been analyzed in \cite{key-5,key-6,key-7,key-8,key-9}.
Resonant production of leptogluons at ep and $\mu$p colliders were
considered in \cite{key-10,key-11,key-12} and \cite{key-13}, respectively.
Indirect production of leptogluons at ILC and CLIC has been studied
in \cite{key-14}. On the other hand, considering IceCube PeV events
\cite{key-15}, color octet neutrinos may be source of these extraordinary
events \cite{key-16}. }

\textcolor{black}{Experimental bound on color octet electron ($e_{8}$),
$M_{e_{8}}>86\:GeV$, presented in \cite{key-17} is based on 25 years
old CDF search for pair production of unit-charged particles which
leave the detector before decaying \cite{key-18}. As mentioned in
\cite{key-19} DO clearly excluded $200\:GeV$ leptogluons decaying
within the detector. The twenty years old H1 search for $e_{8}$  has
excluded the compositeness scale $\Lambda<3\:TeV$ for $M_{e_{8}}\approx100\:GeV$
and $\Lambda<240\:GeV$ for $M_{e_{8}}\approx250\:GeV$ \cite{key-20,key-21}.
While the LEP experiments did not perform dedicated search for leptogluons,
low limits for excited lepton masses, namely $103.2\:GeV$ \cite{key-17},
certainly is valid for $\ell_{8}$, too. Finally, reconsideration
of CMS results on leptoquark searches performed in \cite{key-7} leads
to the strongest current limit on the $e_{8}$  mass, $M_{e_{8}}>1.2-1.3\:TeV$. }

\textcolor{black}{The advantage of ep colliders with sufficiently
high center of mass (CM) energies is that $e_{8}$ is produced in
resonance mode. Large Hadron electron Collider \cite{key-22} (LHeC)
is the highest center of mass energy ep collider proposal up to date.
Unfortunately, approved option which assumes $60\:GeV$ energy recovery
linac for electron beam \cite{key-2323}, will not give an opportunity
to cover $e_{8}$ masses above $1.3\:TeV$ \cite{key-11}. For this
reason, ep colliders with higher energies should be considered for
resonant production of $e_{8}$. }

\textcolor{black}{In this paper, we consider resonant production of
$e_{8}$ at the FCC based ep colliders. Main parameters of these colliders
are given in Section II. Phenomenology of $e_{8}$ is presented in
Section III. In Section IV, signal and background analyses have been
performed and discovery limits on $e_{8}$ mass are estimated. Compositeness
scale matters are discussed in Section V. Finally we summarize our
results in Section VI. }

\section{FCC based $ep$ colliders}

\textcolor{black}{It is widely known that lepton-hadron collisions
have been playing a crucial role in exploration of deep structure
of matter. For example, electron scattering on atomic nuclei reveals
structure of nucleons in Hofstadter experiment \cite{key-23}, quark
parton model was originated from lepton-hadron collisions etc \cite{key-24}.
Investigation of extremely small $x$ but sufficiently high $Q^{2}$
will provide a basis for deeper understanding of the nature of strong
interactions at all levels ranging from nucleus to partons. In addition,
the results from lepton-hadron colliders are necessary for adequate
interpretation of physics at possible future hadron colliders. Today,
linac-ring type ep machines seem to be the most convenient way to
TeV scale in lepton-hadron collisions; and it is also possible that
in future, $\mu$p machines can also be considered depending on solutions
of the principal issues of the $\mu^{+}\mu^{-}$ colliders.}

\textcolor{black}{FCC \cite{key-25} is future $100\:TeV$ CM energy
pp collider proposed at CERN and supported by European Union within
the Horizon 2020 Framework Programme for Research and Innovation.
It includes also an electron-positron collider option at the same
tunnel (TLEP), as well as ep collider options. Construction of future
$e^{+}e^{-}$ colliders and $\mu^{+}\mu^{-}$ colliders tangential
to FCC will give opportunity to achieve highest CM energies in ep
and $\mu$p collisions \cite{key-26,key-27,key-28}. A possible configuration
of FCC based lepton-hadron colliders is shown in Figure 1.}

\textcolor{black}{}
\begin{figure}[H]
\begin{centering}
\textcolor{black}{\includegraphics[scale=0.5]{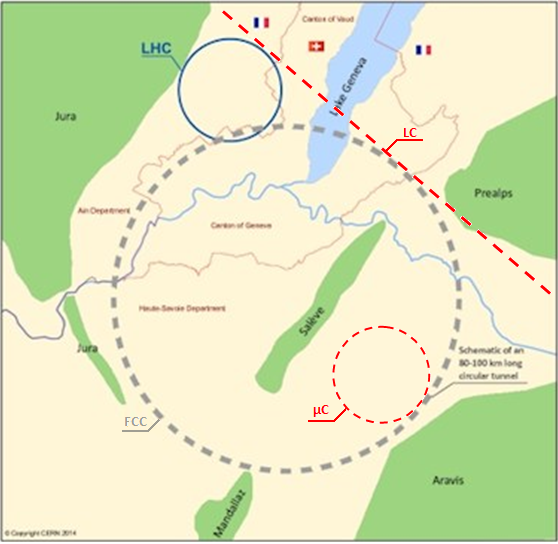}}
\par\end{centering}
\textcolor{black}{\caption{Possible configuratio\textcolor{black}{n of }FCC, linear collider
(LC) and muon collider ($\mu$C).}
}
\end{figure}

\textcolor{black}{CM energy and luminosity values for FCC based ep
colliders - with three different options of electron beam energy -
which we use in this study are given in Table II. In this table ERL60
denotes energy recovery linac proposed for LHeC, ILC means International
Linear Collider \cite{key-29} with highest energy and PWFA-LC denotes
Plasma Wake-Field Accelerator Linear Collider \cite{key-30} with
highest energy (for details see refs. \cite{key-26,key-27,key-28}).
In principle, staged scheme can be considered for the FCC based ep
colliders: starting from ERL60$\varotimes$FCC, through ILC$\varotimes$FCC
to the highest CM energy PWFA-LC$\varotimes$FCC.}

\textcolor{black}{}
\begin{table}[H]
\textcolor{black}{\caption{Main parameters of the FCC based ep colliders.}
}
\centering{}\textcolor{black}{}%
\begin{tabular}{|c|c|c|c|}
\hline 
\textcolor{black}{Collider Name} & \textcolor{black}{$E_{e}$, TeV} & \textcolor{black}{CM Energy, TeV} & \textcolor{black}{$L_{int}$, $fb^{-1}$per year}\tabularnewline
\hline 
\hline 
\textcolor{black}{ERL60$\varotimes$FCC} & \textcolor{black}{0.06} & \textcolor{black}{3.46} & \textcolor{black}{100}\tabularnewline
\hline 
\textcolor{black}{ILC$\varotimes$FCC} & \textcolor{black}{0.5} & \textcolor{black}{10} & \textcolor{black}{10-100}\tabularnewline
\hline 
\textcolor{black}{PWFA-LC$\varotimes$FCC} & \textcolor{black}{5} & \textcolor{black}{31.6} & \textcolor{black}{1-10}\tabularnewline
\hline 
\end{tabular}
\end{table}

\section{Color octet electron}

\textcolor{black}{In fermion-scalar models with colored preons, leptons
are bound states of one fermionic color triplet preon and one scalar
color triplet anti-preon }

\textcolor{black}{
\begin{equation}
\ensuremath{\ell}=(F\bar{S})=3\otimes\bar{3}=1\oplus8\label{eq:1}
\end{equation}
 }

\noindent \textcolor{black}{therefore, each SM lepton has one color
octet partner. In three-fermion models with color triplet fermionic
preons the color decomposition is}

\textcolor{black}{
\begin{equation}
\ell=(FFF)=3\otimes3\otimes3=1\oplus8\oplus8\oplus10\label{eq:2}
\end{equation}
and each SM lepton has two color octet and one color decuplet partners.
Concerning the relation between compositeness scale and masses of
leptogluons, two scenarios can be considered: $M_{e_{8}}\approx\varLambda$
(QCD-like scenario) and $M_{e_{8}}\ll\varLambda$ (Higgs-like scenario).
In the second scenario SM-like hierarchy may be realized, namely,
$M_{e_{8}}\ll M_{\mu_{8}}\ll M_{\tau_{8}}\ll\varLambda$. Interaction
lagrangian of $\ell_{8}$ with leptons and gluons can be written as
\cite{key-11,key-17}}

\textcolor{black}{
\begin{equation}
L=\frac{1}{2\Lambda}\sum_{l}\left\{ \bar{\ell_{8}^{\alpha}}g_{s}G_{\mu\nu}^{\alpha}\sigma^{\mu\nu}\left(\eta_{L}\ell_{L}+\eta_{R}\ell_{R}\right)+h.c.\right\} ,\label{eq:3}
\end{equation}
where $g_{s}$ is strong coupling constant, $\varLambda$ denotes
compositeness scale, $G_{\mu\nu}$ is gluon field strength tensor,
$\ell_{L(R)}$ stands for left (right) spinor components of lepton,
$\ell=e,\:\mu,\:\tau$; $\sigma^{\mu\nu}$ is the antisymmetric tensor
($\sigma^{\mu\nu}=\frac{i}{2}\left[\gamma^{\mu},\:\gamma^{\nu}\right]$),
$\eta_{L}(\eta_{R})$ symbolizes chirality factor. Keeping in mind
leptonic chiral invariance ($\eta_{L}\eta_{R}=0$), we take $\eta_{L}=1$
and $\eta_{R}=0$. Decay width of $\ell_{8}$ given by }

\textcolor{black}{
\begin{equation}
\Gamma(\ell_{8}\rightarrow\ell+g)=\frac{\alpha_{s}M_{\ell_{8}}^{3}}{4\Lambda^{2}},\label{eq:4}
\end{equation}
 }

\noindent \textcolor{black}{where $\alpha_{s}=g_{s}/4\pi$. The decay
width of $e_{8}$ is presented in Fig. 2 for $\varLambda=M_{e_{8}}$
and $\varLambda=100$ TeV.}

\textcolor{black}{}
\begin{figure}[H]
\begin{centering}
\textcolor{black}{\includegraphics[scale=0.12]{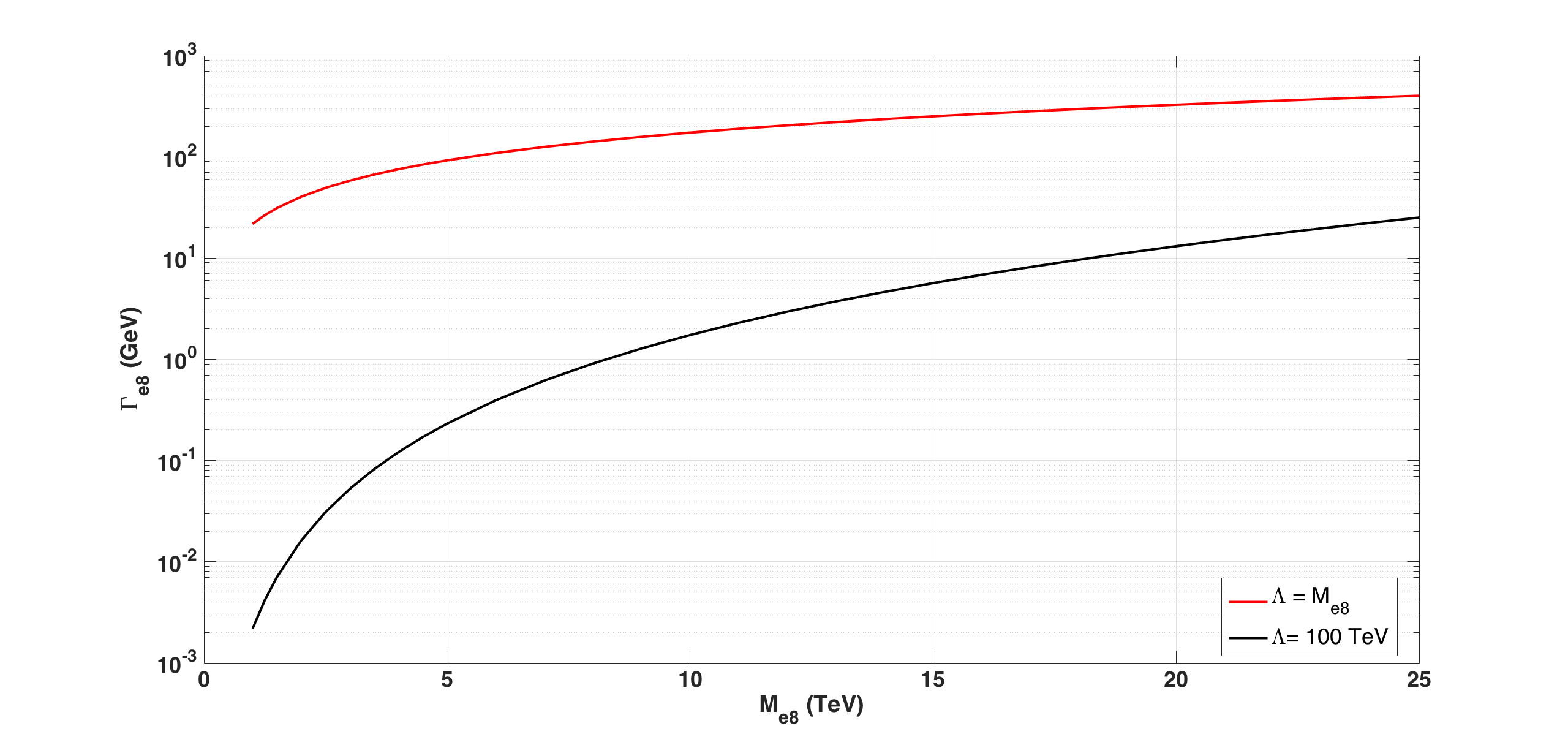}}
\par\end{centering}
\textcolor{black}{\caption{$e_{8}$ decay width vs its mass for $\varLambda=M_{e_{8}}$ and $\varLambda=100$
TeV.}
}

\end{figure}

\textcolor{black}{Diagram for resonant production of $e_{8}$ is shown
in Figure 3. We implement model files of $e_{8}$ into MadGraph5 event
generator \cite{key-31} and use CTEQ6L1 parton distribution function
\cite{key-32} for numerical calculations. MadGraph5-Pythia6 interface
was used for parton showering and hadronization \cite{key-33}. }

\textcolor{black}{}
\begin{figure}[H]
\begin{centering}
\textcolor{black}{\includegraphics[scale=0.4]{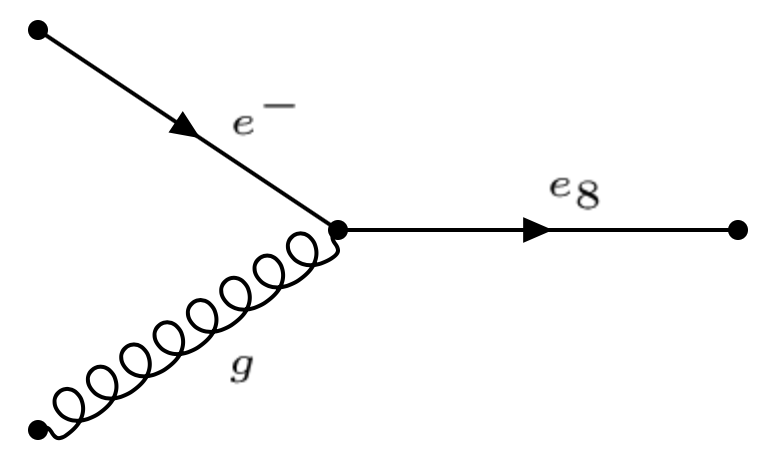}}
\par\end{centering}
\textcolor{black}{\caption{Feynman Diagram for resonant production of $e_{8}$ in ep collisions.}
}

\end{figure}

\textcolor{black}{The resonant $e_{8}$ production cross sections
for different options of the FCC based ep colliders (Table II) are
presented in Fig. 4 (for $\varLambda=M_{e_{8}}$ and $\varLambda=100$
TeV cases). }

\textcolor{black}{}
\begin{figure}[H]
\begin{centering}
\textcolor{black}{\includegraphics[scale=0.2]{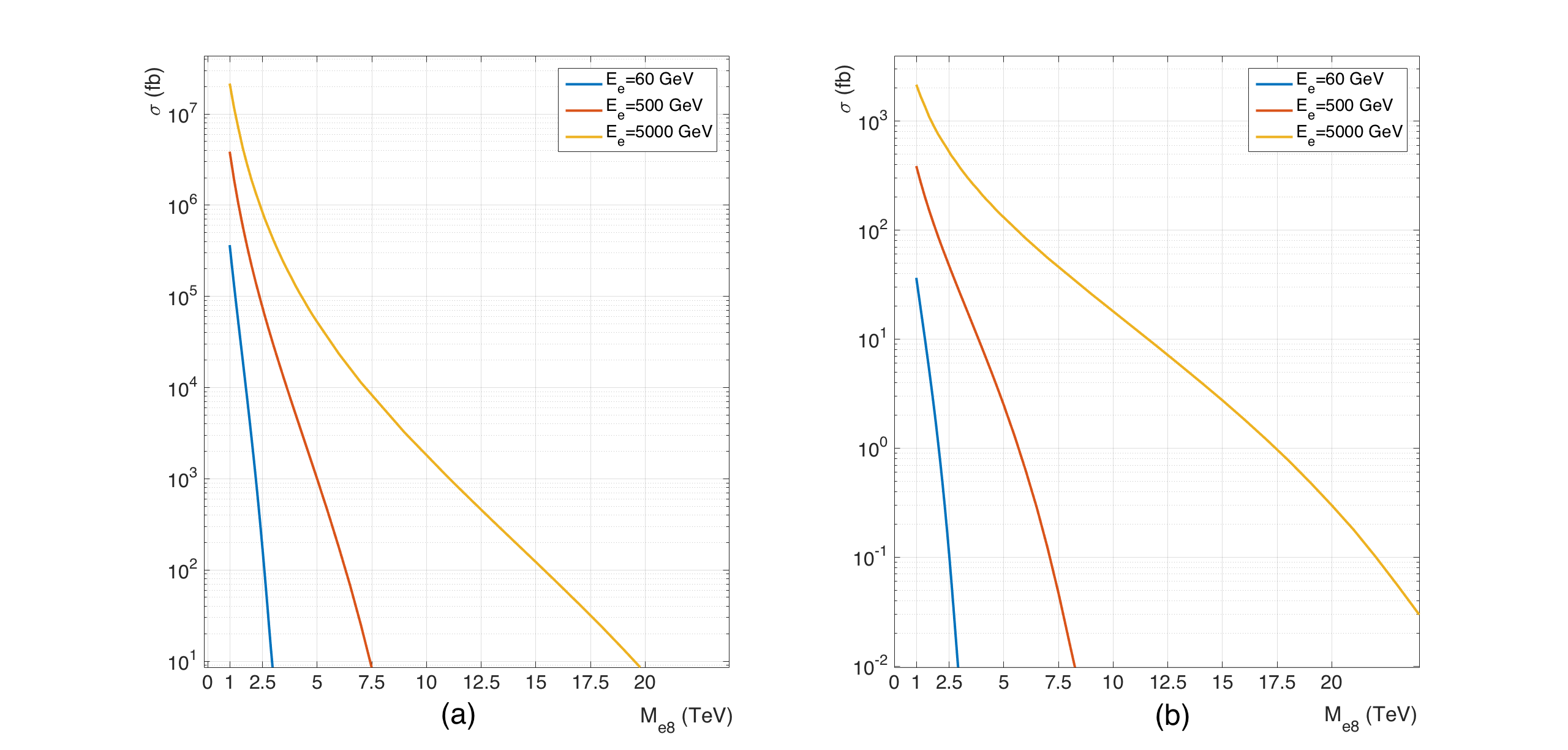}}
\par\end{centering}
\textcolor{black}{\caption{Resonant\textcolor{black}{{} production of} $e_{8}$ at the FCC based
ep colliders for $\varLambda=M_{e_{8}}$ (a) and $\varLambda=100$
TeV (b). }
}
\end{figure}

\textcolor{black}{In order to emphasize the advantage of the resonant
production let us compare the production of $e_{8}$ with mass 10
TeV at PWFA-LC$\varotimes$FCC ($\sqrt{s_{ep}}=31.6$ TeV, $L_{int}=10$
$fb^{-1}$) and FCC-pp option ($\sqrt{s_{pp}}=100$ TeV, $L_{int}=500$
$fb^{-1}$). As seen from Fig. 4, production cross section at ep is
2000 $fb$ for $\Lambda=10$ TeV and 20 fb for $\Lambda=100$ TeV,
whereas corresponding cross sections for pair production of $e_{8}$
at the FCC-pp are $\sim0.50$ $fb$ and 0.29 $fb$, respectively.
Therefore, numbers of produced $e_{8}$ are $n=$20000 at ep and $n=250$
at pp if $\varLambda=10$ TeV. Corresponding numbers for $\varLambda=100$
TeV are $n=200$ and $n=188$, respectively. Keeping in mind that
ep collisions have more clear experimental environment }than pp collisions,
PWFA-LC$\varotimes$FCC seems to be more advantageous even for $\varLambda=$
100 TeV case. 

\section{Signal - Background Analysis}

In this section numerical calculations will be performed for $\varLambda=M_{e_{8}}$
. In order to determine appropriate kinematical cuts $p_{T}$ and
$\eta$ distributions for signal and background processes are computed.
At this stage, generic cuts on electron and jet transverse momentum
are chosen as $p_{T_{e}}=20$ GeV and $p_{T_{j}}=30$ GeV, respectively.
Let us mention that jet corresponds to gluon for signal ($eg\rightarrow e_{8}\rightarrow eg$
at partonic level) and quarks for main background ($eq\rightarrow eq$
through $\gamma$ and $Z$ exchanges) processes. Then discovery cuts
on $p_{T}$ and $\eta$ are determined for different electron beam
energy values and the invariant mass distributions are presented with
these cuts. Finally, discovery limits on the color octet electron
are presented. 

Let us start with ERL60$\varotimes$FCC. Transverse momentum distribution
of final state electrons (the same as for jets) is presented in Figure
5a. Keeping in mind that $M_{e_{8}}<1.3$ TeV is excluded by the reconsideration
of CMS results on leptoquark search {[}6{]}, discovery cut $p_{T}>500$
GeV seems to be adequate. Pseudorapidity distributions for electron
and jets are shown in Figures 5b and 5c, respectively. As can be seen
from Figure 5b, $\eta_{e}>$ 0.5 cut drastically reduces the background
while keeping the signal almost unaffected. In similar manner, $\eta_{j}>2.1$
is chosen. Upper limit for both $\eta_{e}$ and $\eta_{j}$ is taken
as $\eta_{e},\,\eta_{j}$$<4.74$ which corresponds to $1^{o}$ in
proton direction. This value can be covered by very forward detector
as in the LHeC case \cite{key-22}. Invariant mass distributions with
generic cuts and discovery cuts are presented in Figures 5d and 6,
respectively. It is seen that after discovery cuts, background goes
down essentially below signal in the invariant mass distribution.

\begin{figure}[H]
\begin{centering}
\includegraphics[scale=0.2]{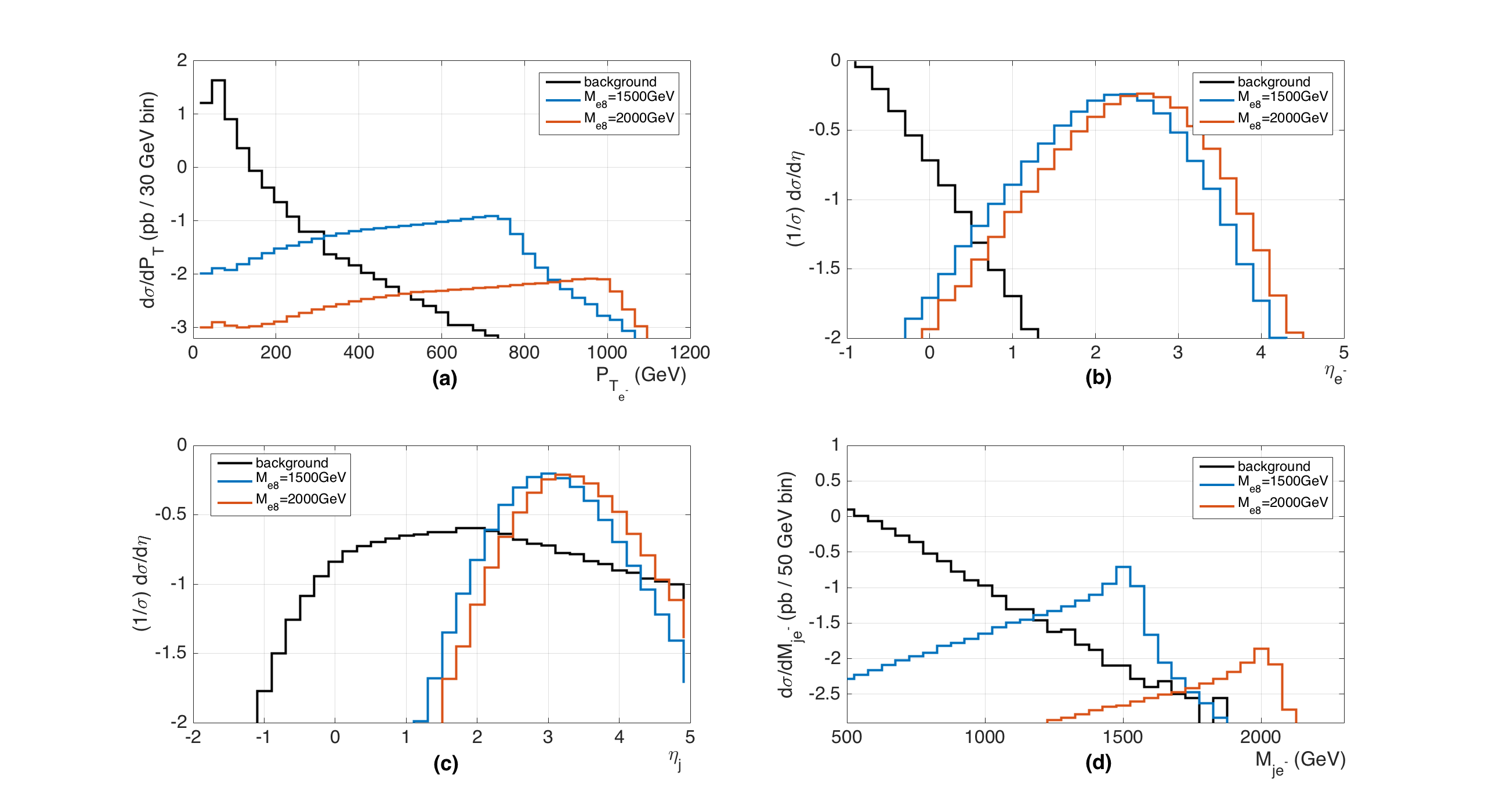}
\par\end{centering}
\caption{a) Transverse momentum distributions of final state jets (and electrons),
b) pseudorapidity distributions of final state electrons, c) pseudorapidity
distributions of final state jets and d) invariant mass distributions
for signal and background at ERL60$\varotimes$FCC after generic cuts. }
\end{figure}

\begin{figure}[H]
\begin{centering}
\includegraphics[scale=0.12]{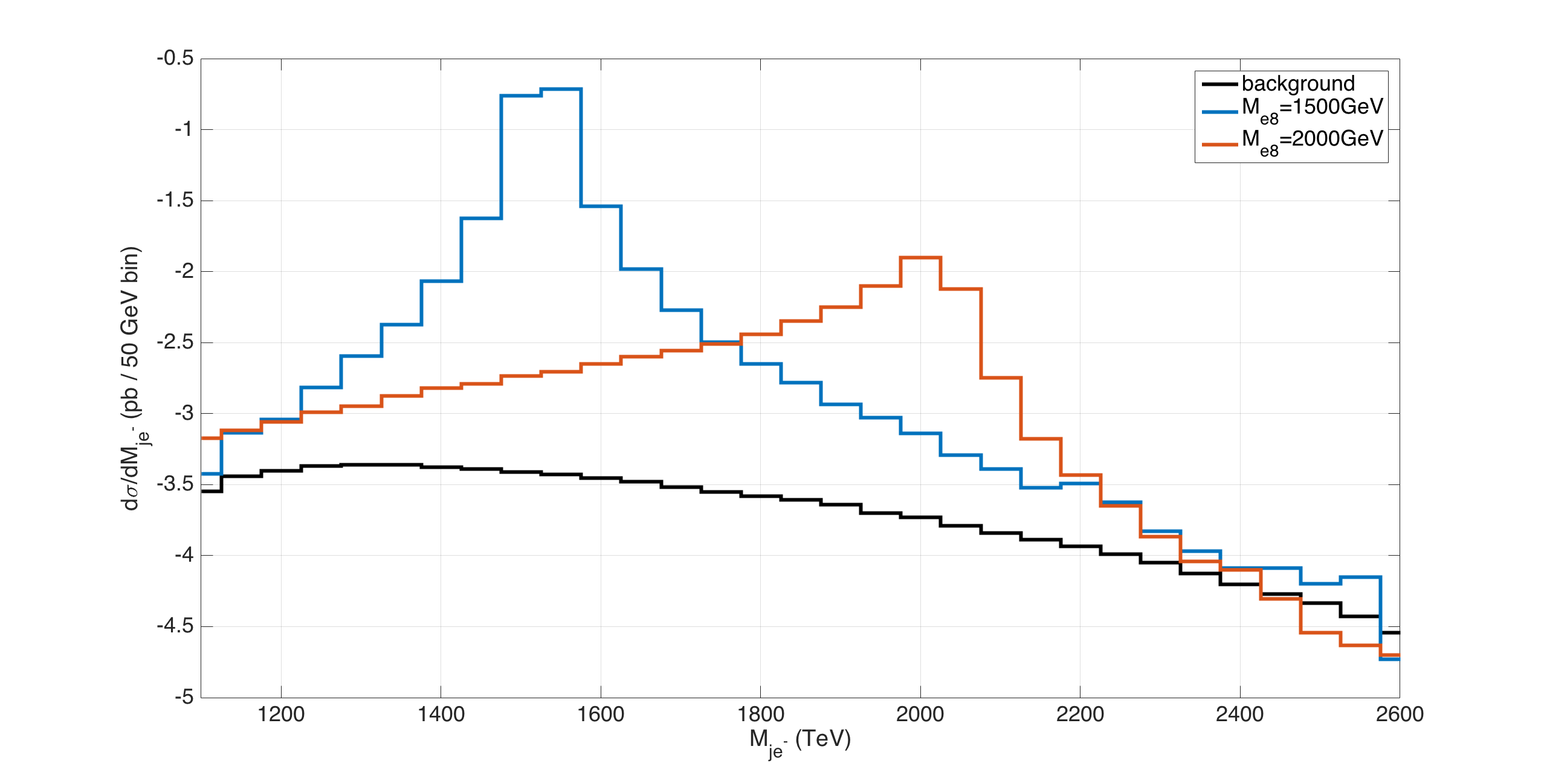}
\par\end{centering}
\caption{Invariant mass distributions for signal and background at ERL60$\varotimes$FCC
after discovery cuts. }
\end{figure}

In order to determine discovery limits for color octet electron, we
use following formula for statistical significance: 

\begin{center}
\begin{equation}
SS=\sqrt{2[(S+B)\,ln(1+(S/B))-S]}\label{eq:5}
\end{equation}
\par\end{center}

\noindent where $S$ and $B$ denote event numbers of signal and background,
respectively. In addition to discovery cuts, mass window cuts are
specified to determine $S$ and $B$ values as $M_{e_{8}}-2\Gamma_{e_{8}}<M_{ej}<M_{e_{8}}+2\Gamma_{e_{8}}$.
Discovery ($SS=5$) and observation ($SS=3$) limits for 100 $fb^{-1}$
integrated luminosity are found to be 2900 and 3100 GeV, respectively.

Performing similar analysis for ILC$\varotimes$FCC and assuming that
\textcolor{black}{$e_{8}$ is not observed by }ERL60$\varotimes$FCC
(that means $M_{e_{8}}>3100$ GeV) we determine following discovery
cuts: $p_{T}>1500$ GeV, $-1.5<\eta_{e}<4.74$ and $0.5<\eta_{j}<4.74$.
Invariant mass distributions after discovery cuts are presented in
Fig. 7. Discovery limits for ILC$\varotimes$FCC with 10 and 100 $fb^{-1}$
integrated luminosities are presented in Table III. Equation 5 and
mass window $M_{e_{8}}-2\Gamma_{e_{8}}<M_{ej}<M_{e_{8}}+2\Gamma_{e_{8}}$
have been used. 

\begin{figure}[H]
\centering{}\includegraphics[scale=0.12]{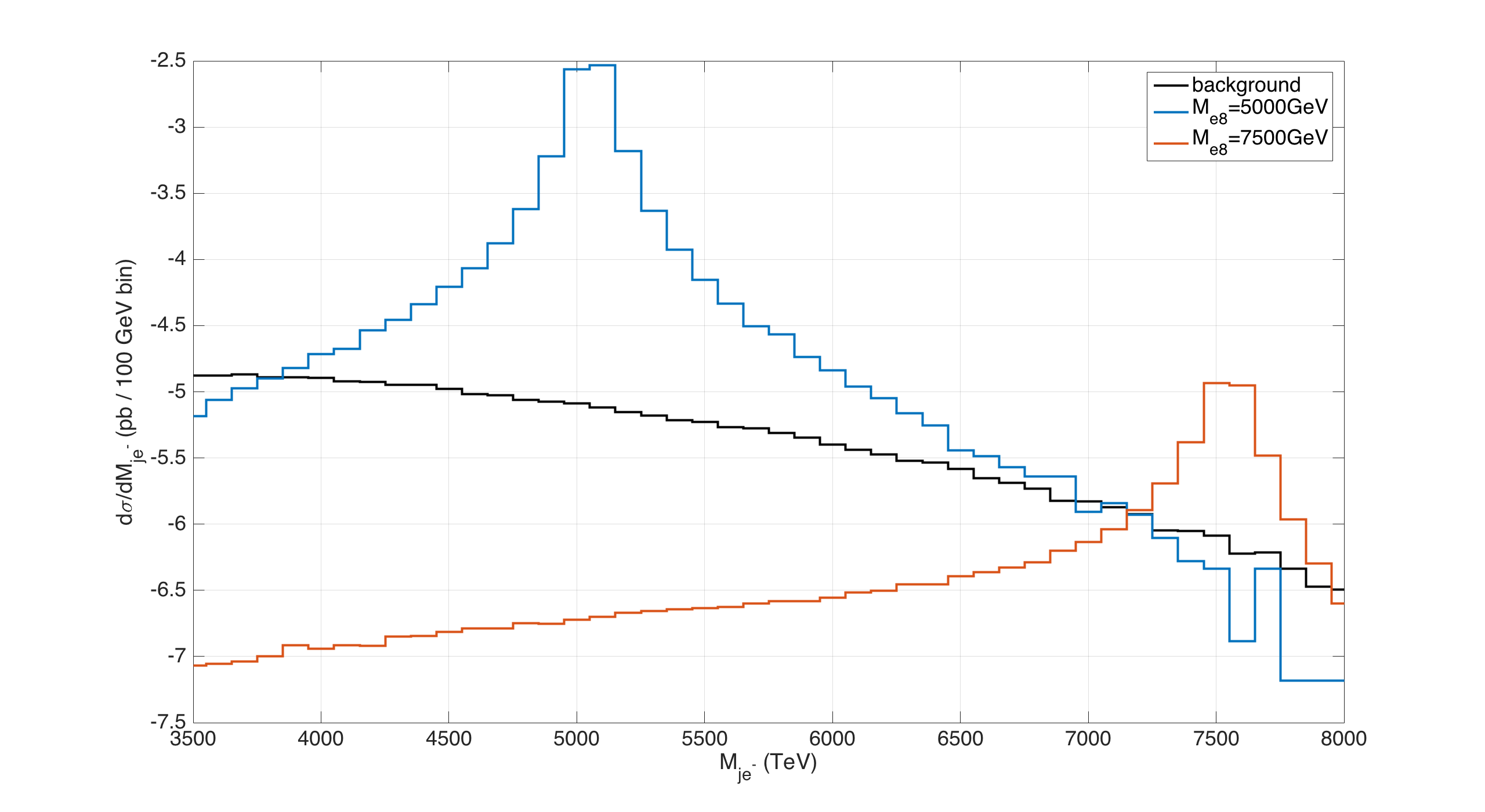}\caption{Invariant mass distributions for signal and background at ILC$\varotimes$FCC
after discovery cuts. }
\end{figure}

\begin{table}
\caption{Observation ($3\sigma$) and discovery ($5\sigma$) limits for $e_{8}$
at different ep colliders.}

\centering{}%
\begin{tabular}{|c|c|c|c|}
\hline 
\multirow{2}{*}{Collider Name} & \multirow{2}{*}{$L_{int}$, $fb^{-1}$ } & \multicolumn{2}{c|}{$M_{e_{8}}$, TeV}\tabularnewline
\cline{3-4} 
 &  & $3\sigma$ & $5\sigma$\tabularnewline
\hline 
\hline 
ERL60$\varotimes$FCC & 100 & 3.1 & 2.9\tabularnewline
\hline 
\multirow{2}{*}{ILC$\varotimes$FCC} & 10 & 8.4 & 8.1\tabularnewline
\cline{2-4} 
 & 100 & 8.9 & 8.6\tabularnewline
\hline 
\multirow{2}{*}{PWFA-LC$\varotimes$FCC} & 1 & 21.6 & 20.1\tabularnewline
\cline{2-4} 
 & 10 & 24.3 & 23.1\tabularnewline
\hline 
\end{tabular}
\end{table}

Similar consideration for PWFA-LC$\varotimes$FCC results in following
discovery cuts: $p_{T}>4000$ GeV (assuming that $e_{8}$ is not observed
at ILC$\varotimes$FCC), $-2.9<\eta_{e}<4.74$ and $-1.0<\eta_{j}<4.74$.
Invariant mass distributions after these cuts are presented in Fig.
8. Discovery limits for PWFA-LC$\varotimes$FCC with 1 and 10 $fb^{-1}$
integrated luminosities are presented in last two rows of Table III.

\begin{figure}[H]
\centering{}\includegraphics[scale=0.12]{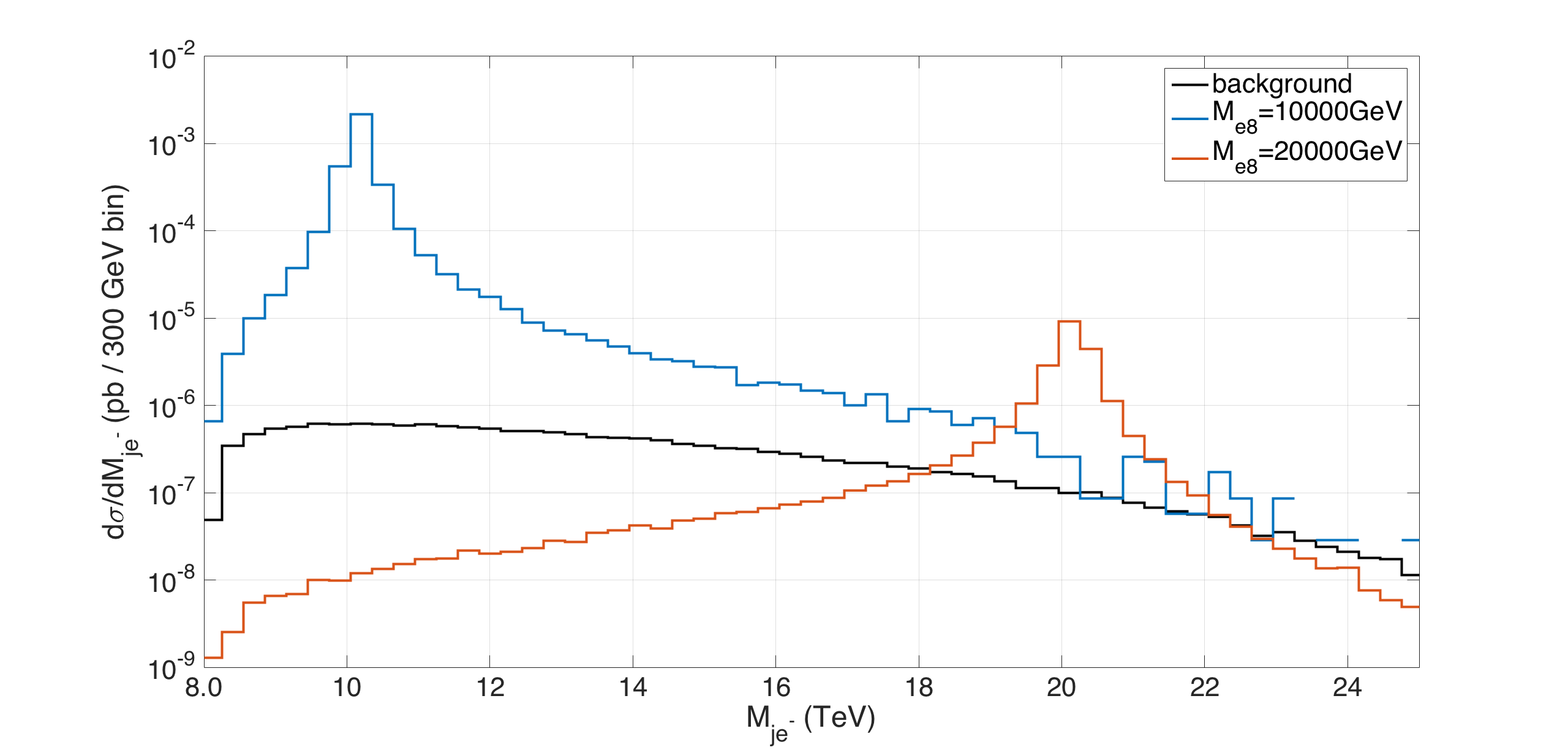}\caption{Invariant mass distributions for signal and background at PWFA-LC$\varotimes$FCC
after discovery cuts. }
\end{figure}

\section{Limits on compositeness scale}

If the $e_{8}$ is discovered by FCC-pp option, ep colliders will
give opportunity to estimate compositeness scale. In this regard,
two distinct possibilities should be considered: 

a)$\;$$e_{8}$ is discovered by FCC but not observed at e-FCC. In
this case one can put lower limit on compositeness scale, 

b)$\;$$e_{8}$ is discovered by FCC and also observed at e-FCC. In
this case one can determine compositeness scale. 

In this section we present the analysis of these two possibilities
for four different benchmark points, namely, $M_{e_{8}}=2.5,\,5,\,7.5$
and $10$ TeV.

\subsection{$e_{8}$ is discovered by FCC but not observed at e-FCC}

Since the $e_{8}$ mass is known one can determine optimal cuts for
given $M_{e_{8}}$. Let us start by consideration of $M_{e_{8}}=2.5$
TeV at ILC$\varotimes$FCC. It is seen from Fig. 5 that $p_{T}>500$
GeV, $-1.30<\eta_{e}<4.74$, $0.50<\eta_{j}<3.00$ cuts drastically
decrease the background whereas the signal is slightly affected. Similar
analyses are performed for other collider options and $M_{e_{8}}$
values. Optimal cuts are presented in Table V. Invariant mass window
$0.99Me_{8}<M_{ej}<1.01Me_{8}$ has been used in this particular analysis.

\begin{table}[H]

\caption{Optimal cuts for determination of compositeness scale lower bounds.}

\centering{}%
\begin{tabular}{|c|c|c|c|c|c|c|c|c|c|}
\hline 
\multirow{2}{*}{Collider} & \multirow{2}{*}{Cut Type} & \multicolumn{2}{c|}{$M_{e_{8}}=2.5$ TeV} & \multicolumn{2}{c|}{$M_{e_{8}}=5.0$ TeV} & \multicolumn{2}{c|}{$M_{e_{8}}=7.5$ TeV} & \multicolumn{2}{c|}{$M_{e_{8}}=10$ TeV}\tabularnewline
\cline{3-10} 
 &  & min & max & min & max & min & max & min & max\tabularnewline
\hline 
\multirow{3}{*}{ERL60$\varotimes$FCC} & $\eta_{e}$ & 0.6 & 4.74 & - & - & - & - & - & -\tabularnewline
\cline{2-10} 
 & $\eta_{j}$ & 2.4 & 4.74 & - & - & - & - & - & -\tabularnewline
\cline{2-10} 
 & Mass Window & 2475 & 2525 & - & - & - & - & - & -\tabularnewline
\hline 
\multirow{3}{*}{ILC$\varotimes$FCC} & $\eta_{e}$ & \textendash 1.3 & 4.74 & -1.1 & 4.74 & -0.8 & 4.74 & - & -\tabularnewline
\cline{2-10} 
 & $\eta_{j}$ & 0.5 & 3.0 & 1.0 & 3.8 & 1.3 & 4.2 & - & -\tabularnewline
\cline{2-10} 
 & Mass Window & 2475 & 2525 & 4950 & 5050 & 7425 & 7575 & - & -\tabularnewline
\hline 
\multirow{3}{*}{PWFA-LC$\varotimes$FCC} & $\eta_{e}$ & -3.3 & 4.74 & -2.9 & 4.74 & -2.7 & 4.74 & -2.6 & 4.74\tabularnewline
\cline{2-10} 
 & $\eta_{j}$ & -1.8 & 0.7 & -1.2 & 1.7 & -0.9 & 2.0 & -0.6 & 2.4\tabularnewline
\cline{2-10} 
 & Mass Window & 2475 & 2525 & 4950 & 5050 & 7425 & 7575 & 9900 & 10100\tabularnewline
\hline 
\end{tabular}
\end{table}

Applying cuts presented in Table V and $p_{T}>500$ GeV for all cases
one can estimate achievable lower limits on compositeness scale. Using
Eq. 5 we obtain $\Lambda$ values given in Table VI. As expected,
lower bounds on compositeness scale is decreased with increasing value
of the $e_{8}$  mass. It is seen that multi-hundred TeV lower bounds
can be put on compositeness scale if $e_{8}$ is discovered at the
FCC and not observed at ILC$\varotimes$FCC and PWFA-LC$\varotimes$FCC.

\begin{table}[H]
\caption{Lower limits on compositeness scale in TeV units at the FCC based
ep colliders}

\centering{}%
\begin{tabular}{|c|c|c|c|c|c|c|c|c|c|}
\hline 
\multirow{2}{*}{Collider} & \multirow{2}{*}{$L_{int}$, $fb^{-1}$} & \multicolumn{2}{c|}{$M_{e_{8}}=2.5$ TeV} & \multicolumn{2}{c|}{$M_{e_{8}}=5.0$ TeV} & \multicolumn{2}{c|}{$M_{e_{8}}=7.5$ TeV} & \multicolumn{2}{c|}{$M_{e_{8}}=10$ TeV}\tabularnewline
\cline{3-10} 
 &  & $3\sigma$ & $5\sigma$ & $3\sigma$ & $5\sigma$ & $3\sigma$ & $5\sigma$ & $3\sigma$ & $5\sigma$\tabularnewline
\hline 
\multirow{1}{*}{ERL60$\varotimes$FCC} & 100 & 44 & 34 & - & - & - & - & - & -\tabularnewline
\hline 
\multirow{2}{*}{ILC$\varotimes$FCC} & 10 & 250 & 195 & 75 & 58 & 22 & 15 & - & -\tabularnewline
\cline{2-10} 
 & 100 & 450 & 350 & 135 & 105 & 42 & 32 & - & -\tabularnewline
\hline 
\multirow{2}{*}{PWFA-LC$\varotimes$FCC} & 1 & 220 & 170 & 200 & 150 & 190 & 145 & 110 & 80\tabularnewline
\cline{2-10} 
 & 10 & 400 & 305 & 390 & 300 & 360 & 275 & 200 & 155\tabularnewline
\hline 
\end{tabular}
\end{table}

\subsection{$e_{8}$ is discovered by FCC and observed at e-FCC}

In this case, the value of cross section at ep colliders which is
inversely proportional to $\varLambda^{2}$ gives opportunity to determine
compositeness scale directly. As an example, let us consider ILC$\varotimes$FCC
case. In Fig. 9 we present $\varLambda$ dependence of $e_{8}$ production
cross section for $M_{e_{8}}=2.5,\,5,\,7.5$ TeV. Supposing that FCC
discovers $e_{8}$ with 5 TeV mass and e-FCC measure cross section
as $\sigma_{exp}\sim2.50$ $fb$, one can derive compositeness scale
as $\varLambda_{exp}=100$ TeV.

\begin{figure}[H]

\begin{centering}
\includegraphics[scale=0.12]{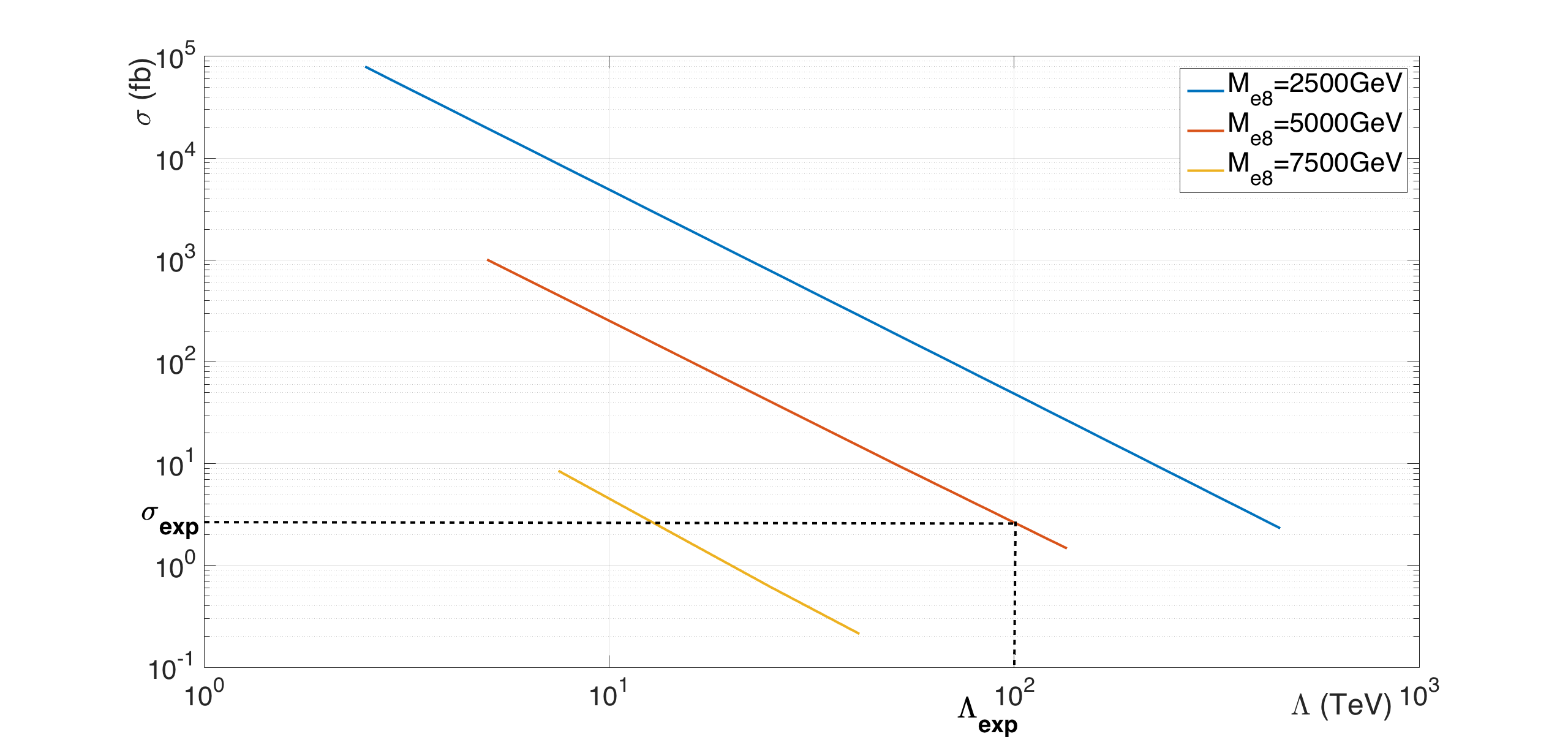}\caption{Cross section distributions with respect to compositeness scale for
ILC$\varotimes$FCC collider. }
\par\end{centering}
\end{figure}

\section{Conclusion}

It seems that FCC based ep colliders have great potential for $e_{8}$
 searches. Discovery limits for $e_{8}$ at the LHC, FCC, ILC, PWFA-LC
and FCC based ep colliders assuming $\Lambda=Me_{8}$ are summarized
in Figure 10. Discovery limit 2.5 TeV for LHC is taken from \cite{key-6}.
A discovery limit of 15 TeV for FCC is obtained by rescaling LHC limit
using the procedure developed by G. Salam and A. Weiler \cite{key-35}.
It is clear that discovery limits for pair production of $e_{8}$
at lepton colliders are approximately $\sqrt{s}/2$. The search potential
of ILC$\varotimes$FCC essentially exceeds that of LHC and linear
colliders whereas is lower than FCC. Highest potential for $e_{8}$
search will be provided by PWFA-LC$\varotimes$FCC with discovery
limit of 23 TeV which is higher than 15 TeV discovery limit provided
by FCC pp collider. On the other hand, observation of $e_{8}$  at
the FCC based ep colliders will provide an opportunity to determine
compositeness scale, in some cases up to multi-hundred TeV. In addition,
polarized e-beams will give opportunity to clarify Lorentz structure
of $e_{8}-e-g$ vertex (this subject is under consideration). 

\begin{figure}[H]
\centering{}\includegraphics[scale=0.15]{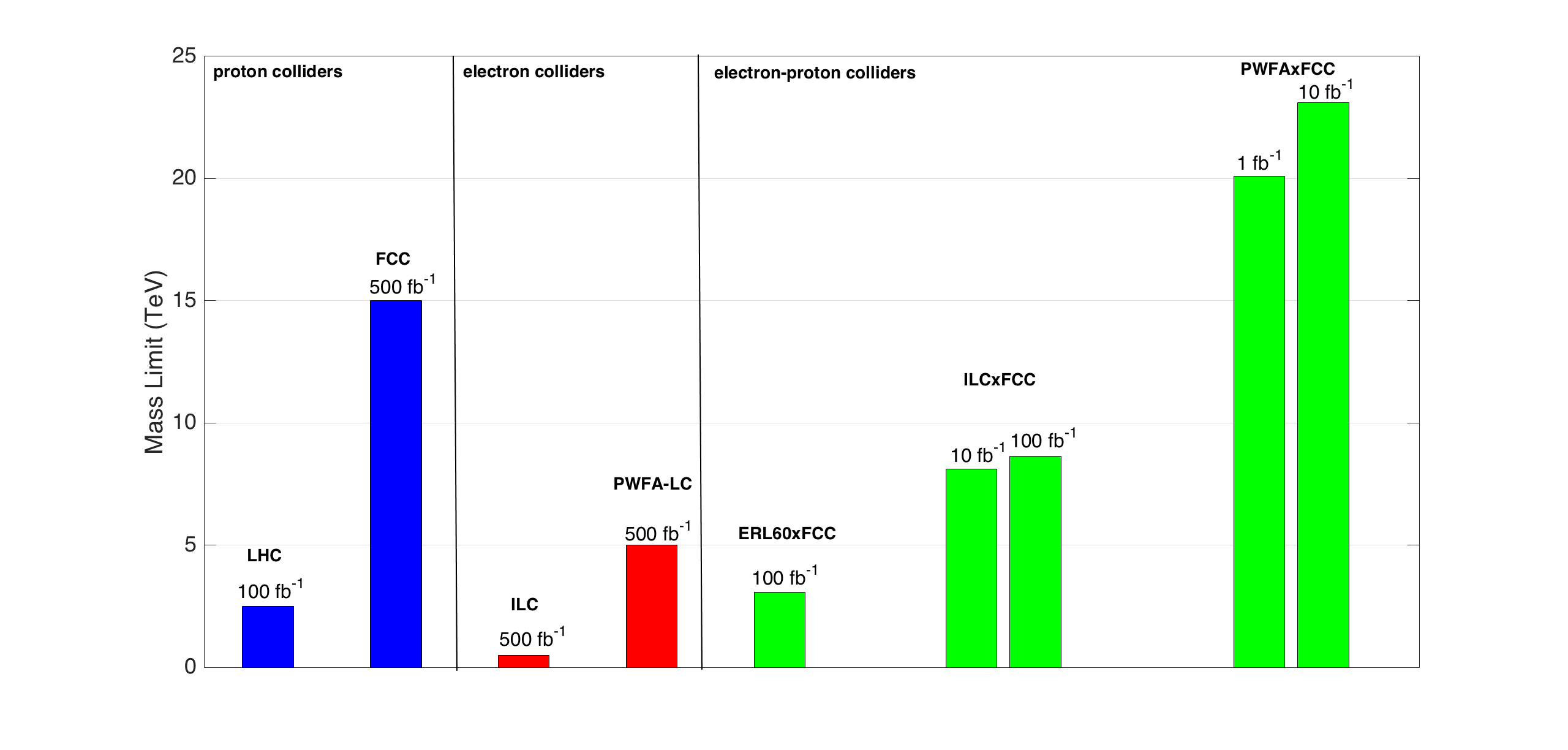}\caption{Discovery limits for $e_{8}$  at different pp, $e^{+}e^{-}$ and
ep colliders.}
\end{figure}

Finally, FCC based energy frontier ep colliders have great potential
for BSM phenomena search, especially when related to the first SM
family fermions. A similar statement is correct for FCC based $\mu$p
colliders if BSM phenomena are related to the second SM family fermions.
Therefore, ERL60$\varotimes$FCC should not be considered as the sole
choise for the FCC based $\ell$p colliders. Energy frontier $\ell$p
options should also be investigated at the same level. The proper
choice for FCC based $\ell$p collider option will be determined by
the FCC results.
\begin{acknowledgments}
This study is supported by TUBITAK under the grant no 114F337. Authors
are grateful to Gokhan Unel and Frank Zimmermann for useful discussions.
Authors are also grateful to Subhadip Mitra and Tanumoy Mandal for
sharing leptogluon MadGraph model file.
\end{acknowledgments}

\end{document}